\documentclass{aa}

\usepackage{epsfig}
\usepackage{graphicx}
\usepackage{natbib}
\usepackage{amssymb}

\newcommand{\bit}{\begin{itemize}}
\newcommand{\eit}{\end{itemize}}
\newcommand{\ben}{\begin{enumerate}}
\newcommand{\een}{\end{enumerate}}
\newcommand{\bfi}{\begin{figure}}
\newcommand{\efi}{\end{figure}}
\newcommand{\beq}{\begin{equation}}
\newcommand{\eeq}{\end{equation}}

\newcommand{\ccu}{cm$^{-3}$ }
\newcommand{\cca}{cm$^{-2}$ }
\newcommand{\ccup}{cm$^{-3}$}
\newcommand{\ccap}{cm$^{-2}$}
\newcommand{\ump}{$\mu$m}
\newcommand{\um}{$\mu$m }
\newcommand{\kmsp}{km s$^{-1}$}
\newcommand{\kms}{km s$^{-1}$ }
\newcommand{\hi}{{\sc Hi} }
\newcommand{\hip}{{\sc Hi}}

\newcommand{\shi}{\mbox{\sc\scriptsize Hi}}

\newcommand{\AaAS}{A\&As}
\newcommand{\AaA}{A\&A}
\newcommand{\ApJ}{ApJ}

\newcommand{\JFM}{J. Fluid Mech.}
\newcommand{\MNRAS}{MNRAS}

\begin{document}

\title{\bf ISOCAM observations of the Ursa Major cirrus:}
\subtitle{Evidence for large abundance variations of small dust grains}
\author{\bf M.-A. Miville-Desch\^enes\inst{1,}\inst{2,}\inst{3}, F. Boulanger\inst{1}, 
G. Joncas\inst{2} and E. Falgarone\inst{3}}
 \institute{Institut d'Astrophysique Spatiale, Universit\'e Paris-Sud, B\^at. 121, 91405, Orsay, France
  \and D\'epartement de Physique, Universit\'e Laval, Sainte-Foy, Qu\'ebec, G1K 7P4, Canada
\and Laboratoire de radioastronomie, \'Ecole Normale Sup\'erieure, 24 rue Lhomond, 75005, Paris, France}
 
\offprints{Marc-Antoine Miville-Desch\^enes}
\mail{miville@lra.ens.fr}
\date{Received 7 March 2001 / Accepted 24 July 2001}

\titlerunning{ISOCAM observations of the Ursa Major cirrus}
\authorrunning{Miville-Desch\^enes et al.}

\abstract{
We present mid-IR imaging observations of a
high Galactic latitude cirrus obtained with the ISO camera ISOCAM at 6" angular resolution.
The observations were done 
with two filters LW2 (5-8.5 \um) and LW3 (12-18 \um) 
that measure respectively the aromatic carbon bands and the underlying 
continuum emission from small dust particles.
Three 0.05 square degree images sample atomic and molecular sections in
the Ursa Major cirrus. These images are compared with \hi, CO and IRAS observations. 
In such a cloud transparent to stellar light ($\rm A_V < 0.5 $)
the mid-infrared to 100 \um and the mid-IR emissivity per hydrogen are
related to the abundance and the optical properties of 
small dust particles independently of any modelling of the penetration of the radiation. 
Within the atomic section of the cloud, the comparison
between ISOCAM images and 21 cm interferometric data 
highlights an enhancement of the mid-IR emitters abundance by a factor $\sim 5$ 
in an \hi filament characterized by a large transverse velocity gradient suggestive
of rotation. Furthermore, a drop in the abundance of the same mid-IR emitters is observed 
at the interface between the atomic and molecular cirrus sections.
We propose that these abundance variations of the mid-IR emitters are related 
to the production of small dust particles by grain shattering in energetic 
grain-grain collisions generated by turbulent motions within the
cirrus and inversely by their disappearance due to coagulation on
large grains.
At the \hip-H$_2$ interface we also observe a change in the $I_\nu$(LW2)/$I_\nu$(LW3) ratio by a factor 2.
This color variation indicates that the amplitude of the continuum near 15 \ump, relative to the aromatic bands, 
rises inside the molecular region. It could result from a
modification of the dust size distribution or of the intrinsic optical properties
of the small dust particles. 
}

\maketitle

\section{Introduction}

We know that interstellar dust plays an important role in the physics  
of the interstellar medium. For instance, dust grains heat 
the gas via the photoelectric effect \cite[]{bakes94} and provide the surfaces necessary to the
formation of H$_2$ \cite[]{duley93}. These processes are sensitive to the size distribution, the 
structure and composition of dust grains and, especially, to the abundance of the 
smallest particles. The IRAS images of nearby molecular complexes 
and bright cirrus clouds ($A_V > 1$) have been extensively used to compare the 
spatial distribution of the mid-IR emission
from the smallest stochastically heated grains and that of the far-IR emission from 
larger grains in thermal equilibrium. The color ratio between these two emissions 
has been related to the abundance ratio between small and large dust grains which was
inferred to widely vary from cloud to cloud and within clouds \cite[]{boulanger90}.
The gain in brightness sensitivity and angular resolution provided since then by dedicated observations
carried out with ISOCAM, the camera on board the Infrared Space Observatory (ISO), 
now allows us to carry on these IRAS investigations and in particular to extend them
to the range of  column densities ($N_{\rm H} \sim 5\times10^{20}$ \ccap) where the transition
from atomic to molecular gas takes place.

ISOCAM has provided low resolution spectra ($\lambda / \Delta \lambda \sim 50$)
of the mid-IR interstellar emission over a wide range of environments including 
the diffuse Galactic emission and cirrus clouds. The emission in the 3-15 \um range is 
invariably associated with emission bands considered to be characteristic
of aromatic hydrocarbon particles. To be heated to temperatures
of a few 100 K by single stellar photons these particles have to
be smaller than a few thousands carbon atoms. The exact nature, structure and 
origin of the emitting particles are still a matter of debate.
The abundance variations revealed by IRAS images show that small dust grains
might not be related to the particles released 
by evolved stars and could be formed within the interstellar medium \cite[]{henning99}. 
It is the motivation of this work to shed further light
on the evolution of small dust grains within the diffuse interstellar medium
in relation to gas physical conditions. To do this we have combined 
ISOCAM observations of the Ursa Major cirrus 
cloud with 21 cm, CO and IRAS data. The Ursa Major cirrus has
a moderate opacity to stellar light ($\rm A_V < 0.5 $) and an atomic and molecular
sections traced by \hi and CO emission respectively. The comparison of these data sets allows us
to measure the mid-IR emissivity per hydrogen which, in a
cloud transparent to stellar light, is
related to the abundance and the optical properties of 
small dust particles independently of any modelling of the 
penetration of the dust heating radiation.

\begin{table*}
\begin{center}
\begin{tabular}{lccc}\hline
& field $A$ & field $B$ & field $C$ \\ \hline\hline
Filters & LW2 and LW3 & LW2 and LW3 & LW2 and LW3\\
Right Ascension (J2000) & 9h 45m 24.8s & 9h 35m 7.4s &  9h 30m 45.6s \\
Declination (J2000) & 70$^\circ$ 22' 54.8'' & 70$^\circ$ 23' 15.1'' & 70$^\circ$ 26' 48.4''   \\
Total number of readouts & 490 & 487 & 490 \\
Number of sky positions & 2 $\times$ 32 & 2 $\times$ 32 & 2 $\times$ 32 \\
Number of readouts per sky position & 5 & 5  & 5 \\
Step size between subsequent positions (pixels) & 8 & 8 & 8\\
Integration time (seconds) & 5.04 & 5.04 & 5.04\\
Lens (pixel field of view) & 6'' & 6'' & 6''\\ \hline
\end{tabular}
\caption{\label{mamd_caract_obs_isocam} Description of the ISOCAM observations. The LW2 and
LW3 broad filters correspond respectively to the wavelength ranges [5-8.5] and [12-18] \ump.
The size of the three fields is $\sim 7' \times 26'$.}
\end{center}
\end{table*}

The observations used in this investigation are presented in \S~\ref{observation}.
The structure and the kinematics of the Ursa Major cirrus is described in \S~\ref{structure}.
In \S~\ref{emissivity} we estimate the emissivity per H atom of small dust grains
in the atomic part. Section~\ref{emiss_v} is devoted to the estimate of the abundance
of small dust grains in an \hi filament with a particular kinematic
and \S~\ref{interface} addresses the same question
for the \hip-H$_2$ interface. The relation between the spatial variations of the dust emission
and the gas physical conditions are discussed in \S~\ref{discussion}. The main
results of this analysis are summarized in \S~\ref{conclusion}.

\begin{figure}
\hspace{-1.2cm}
\includegraphics[width=10.8cm]{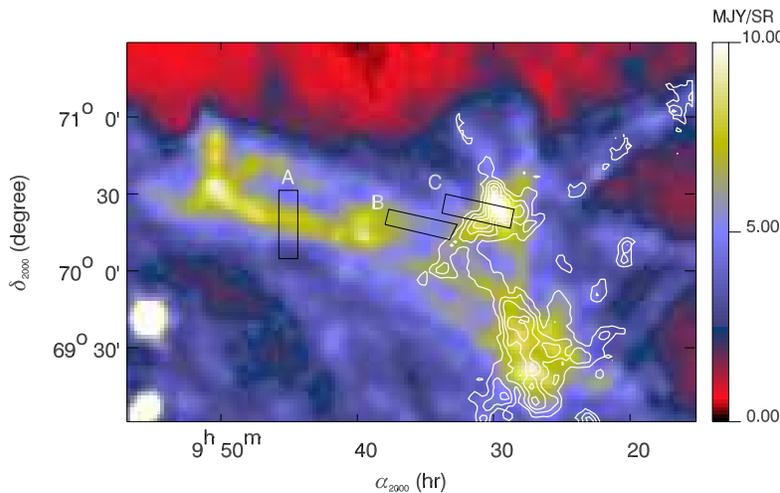}
\caption{\label{mamd_field_position} The Ursa Major cirrus observed by IRAS at 100~\um
(color map) with the CO integrated emission of \cite{pound97} in contours (first level = 5
K \kms and step is also 5 K \kmsp).
The position of the three ISOCAM fields observed are also shown.}
\end{figure}

\section{Observations and data processing}

\label{observation}

We present ISOCAM observations of the Ursa Major cirrus, 
a piece of the North Celestial Loop \cite[]{heiles89}
(see Fig.~\ref{mamd_field_position}). 
This cirrus is composed of two connected parts, one predominantly atomic, 
detected only in the \hi line and one mostly molecular,
detected in CO \cite[]{de_vries87,pound97}. For the sake of simplicity
we will name these two sections of the cirrus {\em atomic} and {\em molecular}
in the following.
The atomic section of the cirrus is representative of
the so called \hi clouds with a density $n\sim 100$ \ccu and a temperature
$T\sim 200$ K \cite[]{joncas92}.
It is illuminated by the mean solar neighbourhood interstellar radiation field (ISRF)
and it is optically thin to this radiation; the \hi column densities range from 
2 to $6\times 10^{20}$ \ccap.
According to \cite{de_vries87}, the distance of the Ursa Major cirrus is $\sim$100 pc.

\subsection{ISOCAM observations}

Three small fields of 0.05 square degree have been mapped with ISOCAM 
(see Table~\ref{mamd_caract_obs_isocam} for the observation log).
The location of the three fields is indicated in Fig.~\ref{mamd_field_position}.
These fields have been selected according to the following criteria.  
First, field $A$ crosses an \hi filament with a large transverse velocity gradient
suggesting a rotation along the long axis
\cite[]{miville-deschenes99b}.
It provides the opportunity of studying the impact of this peculiar
gas
velocity distribution upon the spatial and size distributions of
small dust particles. 
Secondly, fields $B$ and $C$ encompass an atomic-molecular interface,
allowing the analysis of dust properties variations across this interface.

\begin{figure}
\includegraphics[width=\linewidth]{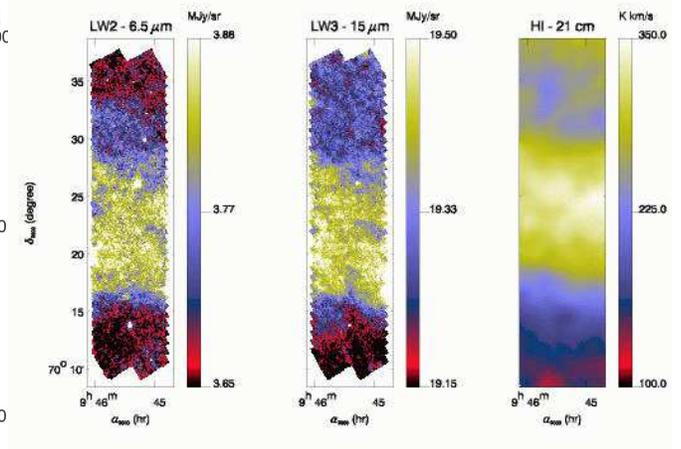}
\caption{\label{mosaicA}
Field $A$ : LW2 image ({\bf left}),
LW3 image ({\bf middle}) and 21 cm integrated emission map 
({\bf right}).}
\end{figure}

\begin{figure*}
\includegraphics[width=\linewidth]{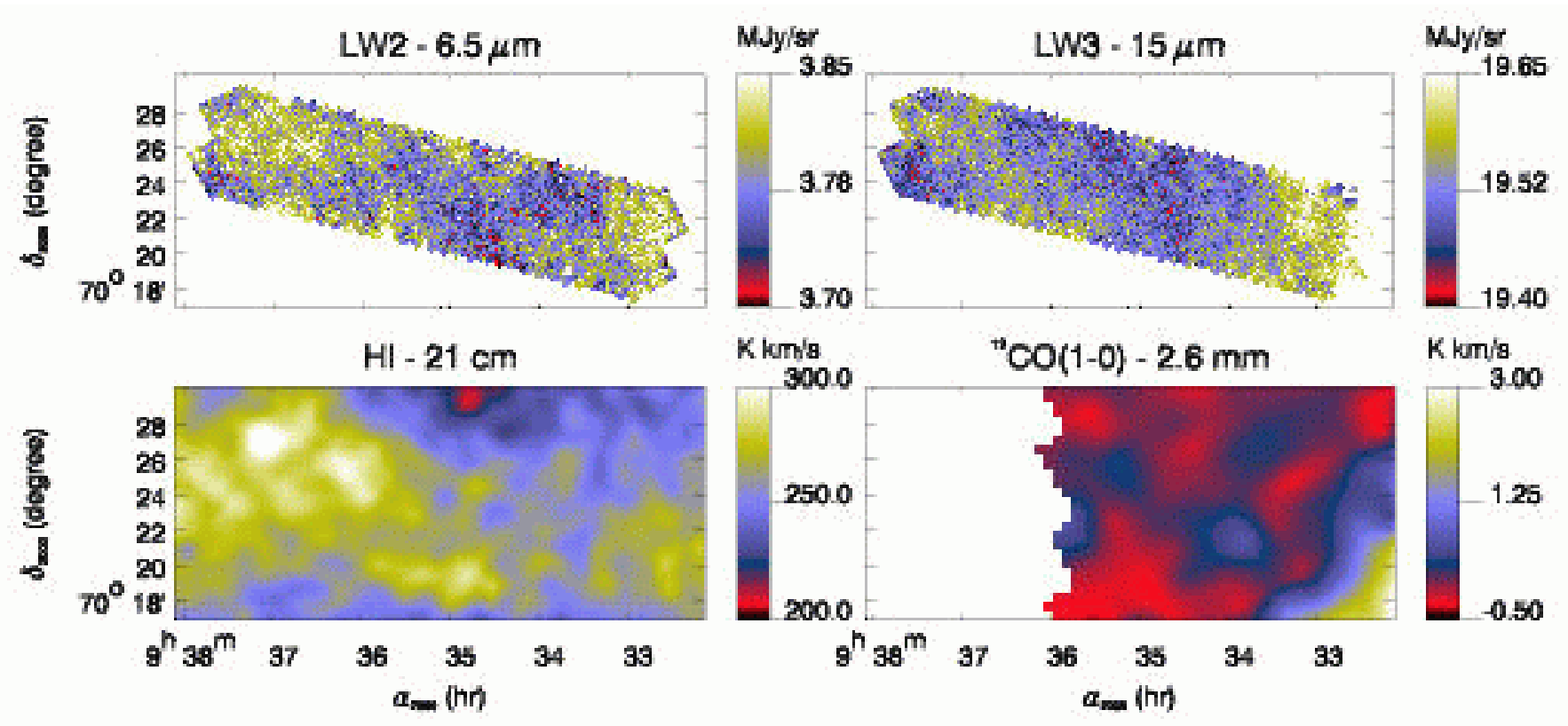}
\caption{\label{mosaicB} ISOCAM (LW2 and LW3), \hi and CO observations of field $B$.} 
\end{figure*}

\begin{figure*}
\includegraphics[width=\linewidth]{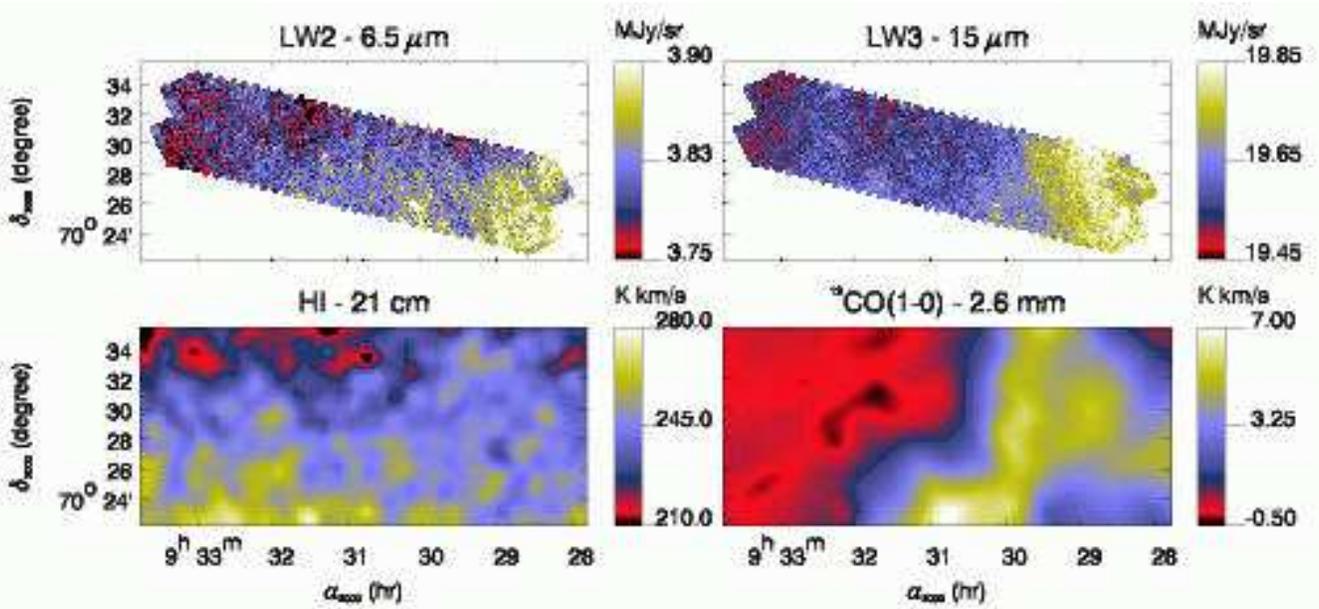}
\caption{\label{mosaicC} ISOCAM (LW2 and LW3), \hi and CO observations of field $C$.} 
\end{figure*}

The ISOCAM observations were made
in the LW2 (5 to 8.5 \ump) and LW3 (12 to 18 \ump) broad band filters.
For diffuse clouds the emission in these filters is dominated by the zodiacal emission; 
the contrast of the cirrus is as small as a few 0.1 per~cent in the LW3 filter.
Furthermore, ISOCAM is affected by instrumental effects that can be one order
of magnitude higher than the contrast we are looking for.
To correct for these instrumental effects we have developed a data processing method 
based on the fact that a given point on the sky has been observed several times.
This method is described in  \cite{miville-deschenes2000a}. Without this particular data processing, 
the ISOCAM observations presented in this paper would have been useless.
The LW2 and LW3 images of the three fields are shown in Figures~\ref{mosaicA}, \ref{mosaicB}
and \ref{mosaicC}. The mid-infrared peak-to-peak fluctuations in these images are $\lesssim 0.4$ MJy sr$^{-1}$ 
and the rms of the high frequency noise is $\sim 0.02$ MJy sr$^{-1}$.

\begin{figure}
\hspace*{-0.7cm}
\includegraphics[width=9.5cm]{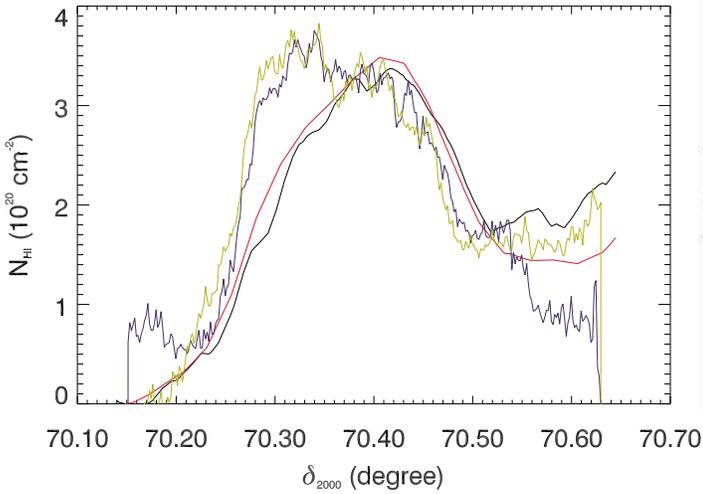}
\caption{\label{cut_LW_HI_IRAS_ISOA} Column density profiles as a function of the declination in field $A$
for the \hi column density (black), LW2 (blue), LW3 (yellow) and 
IRAS 100 \um (red). An offset has been subtracted to each profile (taken at the southern edge of the field).
All the infrared emission profiles were normalized to the \hi column density by constant factors:
$N_{\rm LW2} = (I_{\rm LW2} - 3.61)/6.5 \times 10^{-22}$, $N_{\rm LW3} = (I_{\rm LW3} - 19.12)/9.0 \times 10^{-22}$ and
$N_{100} = (I_{100} - 2.7)/1.4 \times 10^{-20}$}
\end{figure}

\subsection{\hi and CO observations}

We have combined the mid-infrared observations with
CO and \hi observations to conduct our analysis. 
The \hi (21 cm) observations were obtained at the Dominion Radio Astrophysical
Observatory of Penticton (Canada). Part of these data were published by
\cite{joncas92} and the whole set will be the subject of a forthcoming
paper \cite[]{miville-deschenes99b,miville-deschenes2001b}.
The spatial resolution of these 21 cm observations is 1' and the spectral
resolution is 0.66 \kmsp.

In order to compare the dust emission with the whole gas content, we use 
$^{12}$CO observations of the Ursa Major cirrus in the J=(1-0) transition at 115.271203 GHz 
that were obtained by \cite{pound97}. These authors kindly let us use their data
obtained with a Gaussian beam of 100'' and a velocity resolution of 0.27 \kmsp. 
The images of the integrated \hi and CO emission of the three ISOCAM fields
are also shown in Figures~\ref{mosaicA}, \ref{mosaicB} and \ref{mosaicC}.

\section{General structure and kinematics}

\label{structure}

\subsection{Field $A$ - an \hi filament}

\subsubsection{Comparison of dust and gas tracers}

The field $A$ observed by ISOCAM is a small cut across a filament
(length $\sim 5$ pc and  width $\sim 0.5$ pc), 
observed in all IRAS bands (see the 100~\um emission in Fig.~\ref{mamd_field_position}) 
and at 21 cm \cite[]{joncas92}. As seen in Fig.~\ref{mosaicA}, the filament is also
well detected in the mid-infrared. 
The CO observations of \cite{pound97} do not include field $A$, but CO has
been searched unsuccessfully at the 0.1 K level within this area by \cite{de_vries87} and G. Lagache (private
communication). Therefore we will consider that field as predominantly atomic
and use the 21 cm data as a tracer of the spatial structure and kinematic of the gas.

On a large scale, the structure in the LW2 and LW3 broad band filters
is very similar to the \hi emission. 
But a careful comparison of the spatial distribution of the mid-infrared and \hi emissions reveals 
significant discrepancies. In Fig.~\ref{cut_LW_HI_IRAS_ISOA} we plot
the total \hi column density, the LW2, LW3 and IRAS 100 \um
brightness, averaged over $\alpha_{2000}$, 
as a function 
of the declination $\delta_{2000}$ in field $A$. 
In these profiles, the emission in the LW2 and LW3
filters and the 100 \um emission have been normalized to the \hi column density
with a constant ratio. These profiles show that the \hi emission is very 
well correlated to the 100 \um emission, but the peak of the emission in the 
two ISOCAM filters does not coincide with that of the FIR and \hip;
the mid-IR emission is shifted compared to the \hi and 100 \um emissions.

\begin{figure}
\hspace*{-0.5cm}
\includegraphics[width=9.5cm]{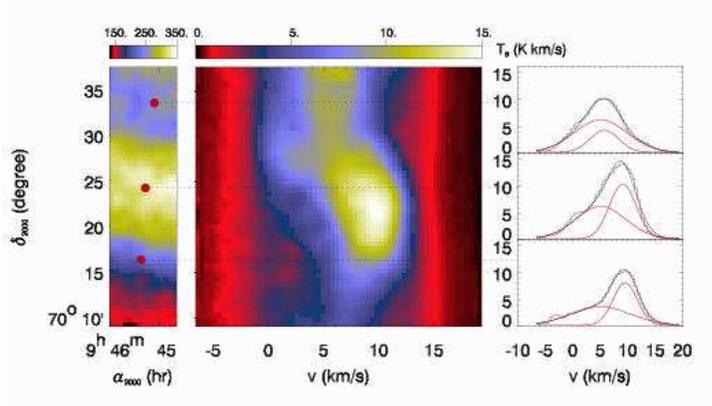}
\caption{\label{posvit} Left: map of the 21 cm integrated flux in field $A$. Center: position 
($\delta_{2000}$)-velocity representation 
of the 21 cm data. Right: three typical spectra with the result of the Gaussian decomposition. The positions
of the three spectra are indicated with red points and dotted lines.}
\end{figure}

\subsubsection{Kinematics of the gas}

\label{kinematicHI}

The velocity structure of the \hi emission within field $A$ is described 
in Fig.~\ref{posvit}.
The central panel presents  a velocity-position diagram
of the 21 cm emission. 
Three representative \hi spectra of this field are also shown in the 
right panel. On most of the field, the 21 cm spectra are characterized by 
two components. The two components
clearly overlap in velocity and it is impossible to separate 
them only by selecting velocity ranges.
To estimate the column density of each component across the field,
we decompose the spectra by a sum of two Gaussians. 
From this decomposition, we derive the integrated intensity, the
velocity centroid and the velocity dispersion of each component on the line of sight. 
The two components  correspond to spatially coherent structures which extend 
beyond the ISOCAM field \cite[]{miville-deschenes2001b}.  The brightest
structure seen in the position-velocity diagram (Fig.~\ref{posvit}) at
a velocity of $\sim 8$ \kmsp corresponds to the filament observed in the 
IRAS 100 \um map (Fig.~\ref{mamd_field_position}). The second component at 
a central velocity of  $\sim 4$ \kmsp is more diffuse;
it is also observed beyond field $A$.

\begin{figure}
\includegraphics[width=\linewidth]{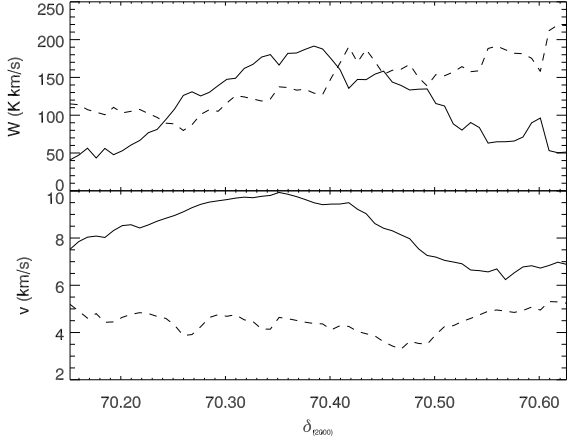}
\caption{\label{coupe_HI_ISOA} Upper panel: Mean integrated 21 cm emission as a function of the declination for the two
\hi components identified by the Gaussian decomposition. Lower panel:
mean velocity as a function of the declination for the two components. The continuous line depicts
the filament and the dotted line depicts the more diffuse component.}
\end{figure}

In Fig.~\ref{coupe_HI_ISOA} we show
the mean integrated intensity and the velocity centroid of these two components
across field $A$.
The component associated with the filament (solid line in Fig.~\ref{coupe_HI_ISOA})
contributes $\sim$ 40\% of the total integrated emission of the field.
It is characterized by two velocity gradients perpendicular to the filament;
first there is a 2 \kms increase of the velocity from $\delta_{2000}=70.15^\circ$ to $70.35^\circ$
and then a decrease of 4 \kms from $\delta_{2000}=70.35^\circ$ to $70.54^\circ$.
At a distance of 100 pc, these variations correspond respectively to
gradients
of $\sim$5 and -10 \kms pc$^{-1}$.
The velocity dispersion of this component measured on individual spectra is $\sim 2-3$ \kmsp.
The more diffuse component (dashed line in Fig.~\ref{coupe_HI_ISOA}), 
which contributes $\sim$ 60\% of the gas emission,
is characterized by a north-south integrated emission gradient and an almost constant velocity 
centroid ($\sim 4$ \kmsp). Its velocity dispersion is significantly larger
than that of the filament, with values between $\sim 6$ and 10 \kmsp.

\subsection{Field $B$ and $C$ - an \hi-H$_2$ interface}

\begin{figure}
\includegraphics[width=\linewidth]{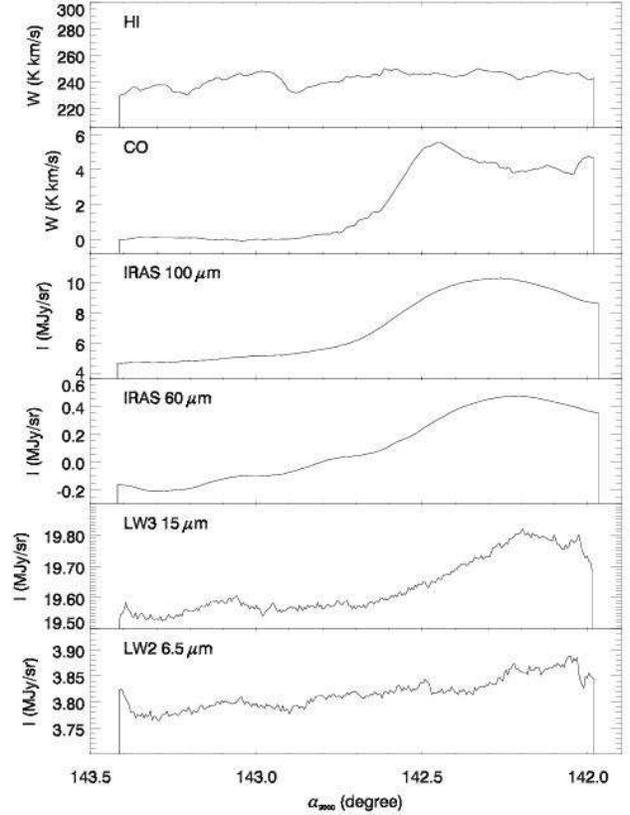}
\caption{\label{coupes_C} Mean intensity in field $C$ as a function of $\alpha_{2000}$.
From top to bottom: \hip (1'), CO (1.7'), IRAS 100 \ump (4'), IRAS 60
\ump (3'), LW3 (6") and LW2 (6"). The numbers in parenthesis are the
resolutions of the observations.}
\end{figure}

The two other fields observed with ISOCAM cross an atomic/molecular
gas interface as indicated by CO observations.
In field $B$ (Fig.~\ref{mosaicB}) there is a slight increase
of the CO emission (up to $T_B \sim 3$ K \kmsp) at the western edge of the field. 
The \hi emission in this field is very uniform; the contrast in the 21 cm line
integrated emission is $\sim$10\%. 

In field $C$, \hi and CO observations are available for the whole region observed by ISOCAM.
There is a slight north-south \hi emission gradient but again the contrast is very low ($< 10$\%).
Most of the column density variations in field $C$ is associated with CO emission.

In Fig.~\ref{coupes_C} we show cuts across field $C$ in \hip, CO, IRAS 100 \ump, IRAS 60 \ump, LW2
and LW3. What is striking at first is the flatness of the \hi emission. There is a sharp rise of the
CO emission near $\alpha_{2000}=142.6^\circ$, not associated with any change in the \hi
emission. In contrast, the dust emission in all 4  bands increases in the molecular
region, but with less sharpness unrelated to the angular resolution.
The infrared dust emission systematically peaks on the western side of
the CO filament. 
Furthermore, we also notice that the peak position of the dust emission at shorter wavelength
seems to be more shifted with respect to the CO emission than at longer wavelength.

\begin{figure}
\includegraphics[width=\linewidth]{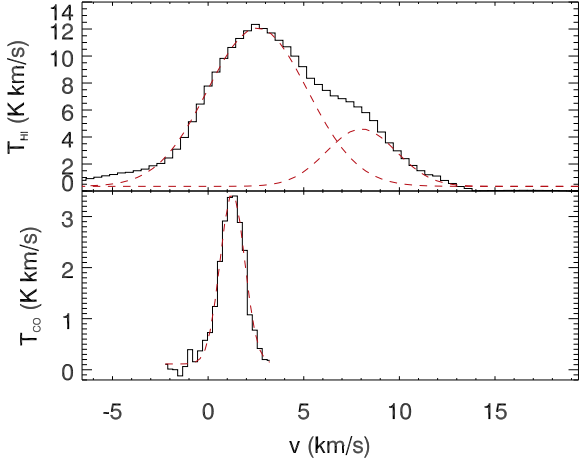}
\caption{\label{spectre_hi_co} {\bf Top:} 21 cm spectrum at the position of the CO J=(1-0) peak intensity in field $C$
($\alpha_{2000}=9h31m$, $\delta_{2000}=70^\circ23'$). {\bf Bottom:} CO spectrum at the same position. The result
of a Gaussian decomposition of the spectra is also shown (dash lines). Two \hi components are
identified (FWHM of 6.16 and 4.28 \kmsp). Only the largest of the two \hi components is seen in CO (FWHM of 1.50 \kmsp).}
\end{figure}

In velocity space, the \hi and CO data are very simple to describe. 
CO and 21 cm spectra taken at the peak CO intensity in field $C$ are shown 
in Fig.~\ref{spectre_hi_co}. These spectra are representative of both fields $B$ and $C$.
The CO emission is characterized by a single symmetric
component centered at $\sim 1.3$ \kms with a velocity width at half power of $\sim 0.7$ \kmsp.
The \hi emission is much broader; the spectrum is clearly not symmetric
with its peak intensity near $\sim 2$ \kmsp. It can be fitted by two Gaussian
components, one at $v \sim 2.7$ \kms with a width of $\sim 2.7$ \kms and a second
one at $v \sim 8.5$ \kms with a width of $\sim 2.0$ \kmsp.

\section{Emissivity of the small dust particles}

\label{emissivity}

\subsection{Correlation between the mid-infrared emission and the \hi integrated emission}

A main goal of this work is to study the relation between the dust and gas emission.
If the dust-to-gas ratio and the heating of the dust are uniform and constant,
the mid-infrared emission $I_{mir}(\alpha, \delta)$ may be described by:
\begin{equation}
\label{emissivity_equation}
I_{mir}(\alpha, \delta) = e \times N_{H} (\alpha, \delta) + I_{back}
\end{equation}
where $e$ is the dust emissivity, $N_{H}$ the total column density of hydrogen (\hi and H$_2$) and 
$I_{back}$ is the background level (assumed constant over the observed
field). The total hydrogen column density 
is deduced from the integrated emission at 21 cm ($W_{\shi}$):
$N_{\shi} = 1.823 \times 10^{18} \, W_{\shi}$ \cite[]{spitzer78} when the atomic gas is
optically thin, which we consider a reasonable assumption for this cloud. 
We did not  look for a correlation between $N_{H}$ and $I_{mir}$ in fields $B$ and $C$
because of the very low contrast of the mid-infrared emission and because of the great uncertainty
concerning the CO to H$_2$ conversion factor (\cite{de_vries87} give 
$N_{H2} = 0.5 \pm 0.3 \times 10^{20}\, W_{CO}$
\cca K$^{-1}$ (km/s)$^{-1}$ for the Ursa Major cirrus). 

\begin{figure}
\hspace*{-0.8cm}
\includegraphics[width=10cm]{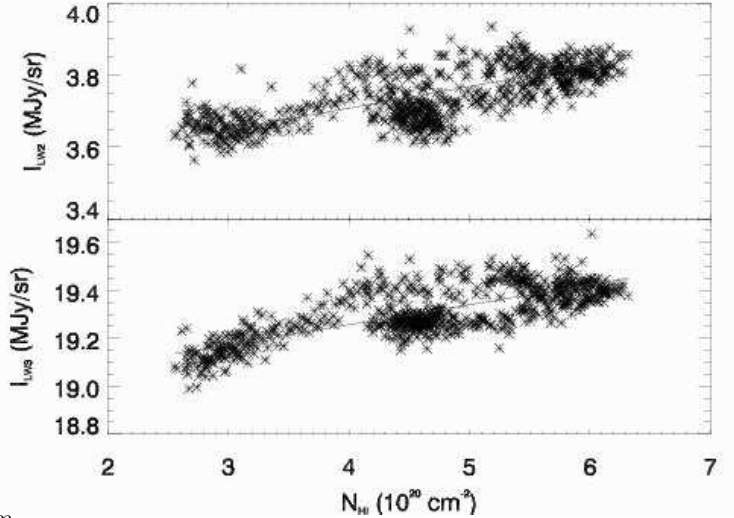}
\caption{\label{LW2_3_hi_relation}
LW2 ({\bf top}) and LW3 ({\bf bottom}) brightness as a function of the \hi column density
(defined in the optically thin approximation). 
The linear regression over-plotted are: $I_{\rm LW2} \mbox{ (MJy/sr)} = 6.07 \times 10^{-22} \, N_{\shi} \mbox{ (cm$^{-2}$)} + 3.47$ and
$I_{\rm LW3} \mbox{ (MJy/sr)} = 8.41 \times 10^{-22} \, N_{\shi} \mbox{ (cm$^{-2}$)} + 18.92$}
\end{figure}

The LW2 vs $N_{\shi}$ and LW3 vs $N_{\shi}$ plots in field $A$ are shown in Fig.~\ref{LW2_3_hi_relation}.
The mid-infrared emissions rise  globally with the \hi column density; these relations are approximately linear
(see Fig.~\ref{LW2_3_hi_relation}). Nevertheless, the rms scatter 
in the LW2-$N_{\shi}$ and LW3-$N_{\shi}$
relations 
(0.05 MJy/sr for LW2 and 0.07 MJy/sr for LW3) is significantly
larger than the noise level (0.02 MJy/sr). 

\subsection{Background emission in the ISOCAM images}

The scatter around the LW2 vs $N_{\shi}$ and LW3 vs $N_{\shi}$ relations indicates that
our description of the mid-infrared emission with Eq.~\ref{emissivity_equation} is probably too simple.
This scatter may originate from a spatial variation of the emissivity or of the background emission.
Here the background level is mainly due to the zodiacal light, which
is uniform at the scale of our small fields.  
Therefore we conclude that the scatter of the LW2 vs $N_{\shi}$ and LW3 vs $N_{\shi}$ relations
in field $A$ is due to spatial variations of the emissivity $e$.

In order to quantify the mid-infrared emissivity for each position ($\alpha$, $\delta$), 
the uniform background emission level of the ISOCAM images must be estimated. 
This is also essential to compare the emission between the LW2 and LW3 filters.
We can estimate the background level by inspecting the relation between the mid-infrared fluxes 
and the total column density of hydrogen (see Fig~\ref{LW2_3_hi_relation}).
If the mid-infrared flux is, at first order, proportional to the total column 
density of hydrogen,
the background level should be the mid-infrared flux for zero column
density of hydrogen. If we apply this to the data
the uncertainty on the background will be large because
the dynamical range of the ISOCAM images is
small  ($\lesssim 0.6$ MJy/sr). Further, in fields $B$ and $C$,
CO emission cannot be considered as a quantitative tracer of the 
H$_2$ column density.
In this context, we chose  
to work in relative brightness for all tracers. This means
that, for each field and for gas and dust maps, 
we subtract an offset taken in a common 
region where the mid-IR emission is the lowest.

\section{Variation of the mid-infrared emissivity with gas velocity}

\begin{figure}
\includegraphics[width=\linewidth]{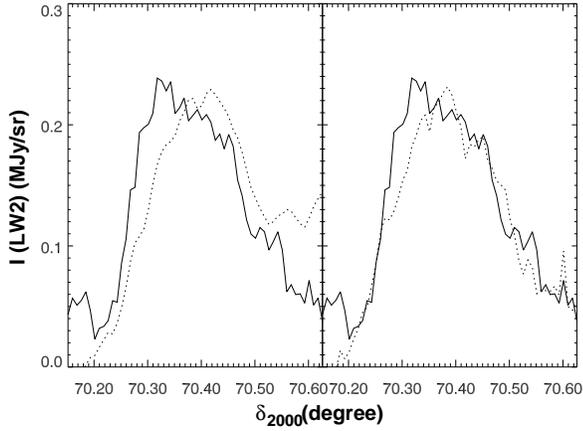}
\caption{\label{coupe_hi_lw2} Mean LW2 flux (at the resolution of the \hi observation) 
as a function of the declination (continuous line). Left: the dotted line
is the mid-infrared profile reproduced from the \hi data with an emissivity of 
$e_\nu = 4.5\times 10^{-22}$ MJy/sr cm$^2$ (which corresponds to
$4\pi \nu e_\nu = 3.8 \times 10^{-31}$ W/H)
Right: the dotted line  is a reconstruction of the mid-infrared profile with an emissivity of $8.2\times 10^{-22}$ MJy/sr cm$^2$
($4.6 \times 10^{-31}$ W/H) for the filament and of $1.6\times 10^{-22}$ MJy/sr cm$^2$ 
($0.9 \times 10^{-31}$ W/H) for the diffuse component.}
\end{figure}

\label{emiss_v}

In this section we show that the spatial shift between the dust and gas tracers in field $A$, and therefore 
the dispersion in the LW2-$N_{\shi}$ and LW3-$N_{\shi}$ relations, 
can be accounted for by a difference in the small dust particles
emissivity per H atom between the two 
\hi kinematic components (see \S~\ref{kinematicHI}).

\begin{figure}
\includegraphics[width=\linewidth]{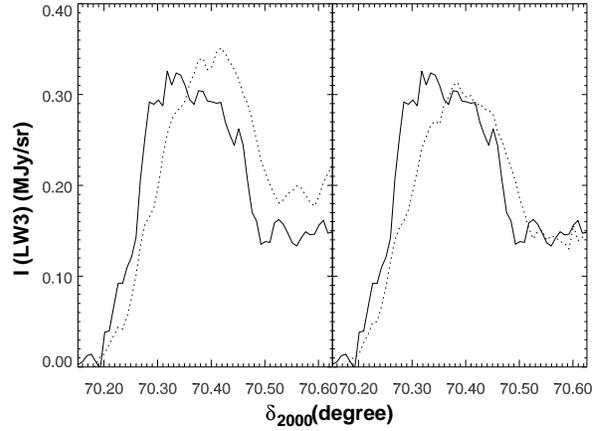}
\caption{\label{coupe_hi_lw3} Mean LW3 flux (at the resolution of the \hi observation) 
as a function of the declination (continuous line). Left: the dotted line
is the mid-infrared profile reproduced from the \hi data with an emissivity of 
$7.1\times 10^{-22}$ MJy/sr cm$^2$ ($2.3 \times 10^{-31}$ W/H).
Right: the dotted line  is a reconstruction of the mid-infrared profile with an emissivity of 
$1.1\times 10^{-21}$ MJy/sr cm$^2$ ($2.5 \times 10^{-31}$ W/H)
for the filament and of $3.8\times 10^{-22}$ MJy/sr cm$^2$ ($1.0
\times 10^{-31}$ W/H) for the diffuse 
component.}
\end{figure}

As seen in Fig.~\ref{coupe_HI_ISOA}, the two \hi components identified 
earlier with the Gaussian decomposition have very
different column density profiles across field $A$. 
If we consider the morphology of both components, the mid-infrared emission is more similar
to the filament than to the diffuse \hi component. By giving different emissivity to these
two \hi components we were
able to reproduce rather well the mid-infrared profiles (see Fig~\ref{coupe_hi_lw2} and \ref{coupe_hi_lw3}). 
The LW2 and LW3 emissivities that best reproduce the mid-infrared observations 
are detailed in Table~\ref{emissivity_table}.

The comparison of the mean emissivity of typical Galactic cirrus of the solar neighborhood 
obtained at 12 \um by \cite{boulanger88} with the LW2 and LW3 emissivities ($4\pi \nu e_\nu$ in W/H) deduced here, 
shows that the diffuse component of field $A$ has an emissivity very similar to the average cirrus value.
On the other hand, the filamentary component seems to have an emissivity $5\pm 1$ (LW2) 
and $2.5\pm 0.5$ (LW3) times higher than
the average value. From the numbers of Table~\ref{emissivity_table}  we also conclude that
the ratio of the LW2 and LW3 emissivity is $\sim 0.9$ 
in the diffuse gas and $\sim 1.8$ in the filament. 
Because of the good spatial correlation between the 100 \um and \hi
column density (see Fig.~\ref{cut_LW_HI_IRAS_ISOA}), the large grains
responsible for the 100 \um emission must have the same emissivity in the
the filamentary and diffuse components.

\begin{table}[!ht]
\begin{center}
\begin{tabular}{lllc}\hline
 & & & $4\pi\nu e_\nu$ \\
 Field & $\lambda$ (\ump) & Component & ($10^{-31}$ W/H) \\\hline
A & 6.5 (LW2) & average &  $3.8 \pm 0.5$ \\
& & filament & $4.6 \pm 0.3$\\
& & diffuse gas & $0.9 \pm 0.3$ \\ \cline{2-4}

 & 15.0 (LW3) & average & $2.3 \pm 0.1$\\
& & filament & $2.5 \pm 0.3$\\
& &  diffuse gas & $1.0 \pm 0.3$\\  \hline \hline

C & 6.5 (LW2) & \hi &  $1.9 \pm 0.2$ \\
& & CO & $0.48 \pm 0.05$\\ \cline{2-4}

& 15.0 (LW3) & \hi &  $1.3 \pm 0.1 $ \\
& & CO & $0.61 \pm 0.06$\\  \hline \hline

high latitude & 12.0 & & $1.1 \pm 0.3$\\\cline{2-4}
 & 100.0 & & $3.2 \pm 0.1$\\ \hline

\end{tabular}
\caption{\label{emissivity_table} Emissivity in LW2 and LW3 in fields $A$ and $C$.
The 12 and 100 \um emissivity of typical high latitude Galactic cirrus \cite[]{boulanger88} are also given
as a comparison. The uncertainties on the emissivity for the diffuse and filamentary components 
of field $A$ reflect the uncertainties we have on the determination of the background emission.
The uncertainty on the average emissivity of field $A$ has been computed using the bootstrap method \cite[]{efron86}
on the LW2-N$_{\shi}$ and LW3-N$_{\shi}$ relations of Fig.~\ref{LW2_3_hi_relation}.
The uncertainty on the emissivities of field $C$ reflects the variations of the emissivity values
over the region where they were measured.}
\end{center}
\end{table}

\section{Abundance variations of small dust grains in an  \hip-H$_2$ interface}

\label{interface}

\begin{figure}
\includegraphics[width=\linewidth]{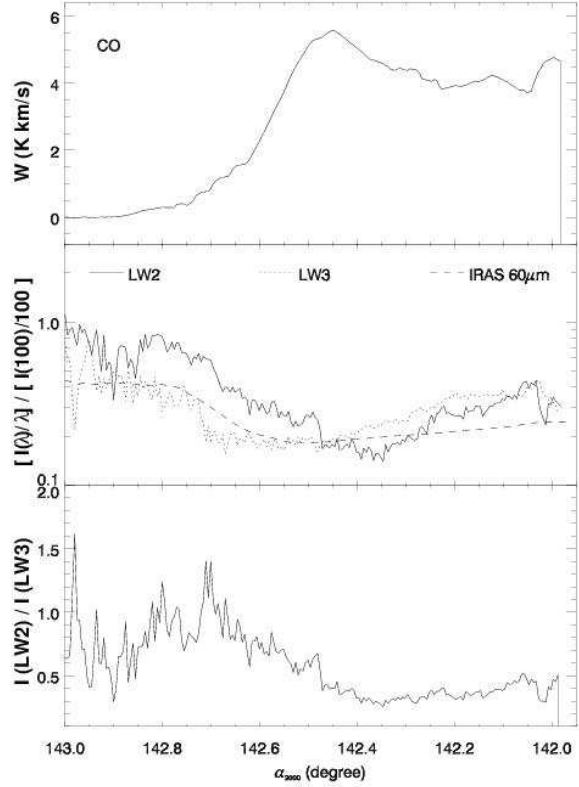}
\caption{\label{emiss_CO_C} {\bf Top:} Mean CO emission of the western part of field $C$, as a function of $\alpha_{2000}$.
{\bf Middle:} Ratio $(I_{lw2}/6.5)/(I_{100}/100)$, $(I_{lw3}/15)/(I_{100}/100)$ and $(I_{60}/60)/(I_{100}/100)$ 
as a function of $\alpha_{2000}$. {\bf Bottom:} Ratio of the LW2 over LW3 emission as a function of $\alpha_{2000}$.}
\end{figure}

Fields $B$ and $C$ cross an interface between a mainly atomic region and another one where
significant CO emission is detected. The emission in field $B$ is very uniform;
it is thus difficult to correlate variations in the dust and gas emissions.
Therefore, our whole analysis of the \hip-H$_2$ transition will be based on the observations of field $C$.

To estimate the dust mid-infrared emissivity, we usually rely on an estimate of the gas column density.
In field $C$, the structure of the gas is dominated by the molecular component. Because the
determination of the CO to H$_2$ ratio is poorly constrained and because, in such diffuse clouds, it is
known that CO is not such a good tracer of the molecular column density \cite[]{reach94},  
we prefer to compare the mid-infrared emission with the IRAS 100 \um band emission to estimate 
the abundance of the smallest dust particles, relative to that of bigger grains responsible for the emission
at 100 \ump.

In Fig.~\ref{emiss_CO_C} (middle panel) we show the ratio
$(\nu I_\nu)_\lambda/(\nu I_\nu)_{(100 \mu m)}$
for LW2 ($\lambda$=6.5 \ump),  LW3 ($\lambda$=15 \ump) and IRAS 60 \ump, across the \hip-H$_2$ transition.
This way of presenting the data emphasizes the variation of dust emissivity at wavelength $\lambda$
compared to the emissivity at 100 \ump. If the grain properties and the heating are homogeneous, this
ratio reflects changes in the relative abundance of grains. For all three wavelengths a decrease
in the $(\nu I_\nu)_\lambda/(\nu I_\nu)_{(100 \mu m)}$ ratio is observed, correlated with the increase
of the CO emission. Furthermore, for the three bands, the lowest $(\nu I_\nu)_\lambda/(\nu I_\nu)_{(100 \mu m)}$
is observed near the CO peak emission.

By making the assumption that the 100 \um emissivity is that measured by \cite{boulanger88}
for the high latitude clouds (see Table~\ref{emissivity_table}), which
is in accordance with the 100 \um emissivity estimated in that cloud
by \cite{joncas92}, we were able to estimate the
emissivity in the LW2 and LW3 bands from the $(\nu I_\nu)_\lambda/(\nu I_\nu)_{(100 \mu m)}$ ratio.
In Table~\ref{emissivity_table} we list an estimate of the mean LW2 and LW3 emissivity values for the atomic 
and molecular regions of the \hip-H$_2$ transition. We estimate that the ratio of the emissivity
in the \hi region compared to the CO region is $4.0 \pm 0.8$ and $2.1 \pm 0.4$ in LW2 and LW3
respectively.

The ratio of the LW2 over LW3 emission for field $C$ is also shown in Fig~\ref{emiss_CO_C}. 
In the purely atomic part of the fields, the ratio 
is $\sim 0.8$ and it drops to $\sim 0.3$, in a few 0.1pc, as we enter the molecular region.

\section{Discussion}

The comparison of the mid-infrared, \hip, CO and IRAS observations 
has allowed us to highlight important spatial variations
of the small dust particles emissivity in the Ursa Major cirrus. 

\label{discussion}

\subsection{Spatial variation of the mid-infrared emissivity}

The Ursa Major cirrus is characterized by a low gas column density
($6.0 \times 10^{20}$ \ccap - corresponding to $A_V \approx 0.3$)
and the absence of star formation. In this context, the spatial variations in 
the mid-infrared emissivity can result from either (1) the 
presence of molecular gas or very cold \hi gas
or (2) an attenuation of the UV radiation field on scales smaller than the beam size
or (3) variations in the small grains abundance or emission properties.
The good correlation observed between the \hi and $100 \mu m$ emissions 
in Fig.~\ref{cut_LW_HI_IRAS_ISOA} puts constraints on hypothesis (1) and (2).
Dust associated with molecular gas should produce an excess of IR emission with respect to the
IR-\hi correlation. However, this excess could be reduced by an attenuation
of the UV radiation field in the same regions since molecules are expected to be present
in the most shielded parts of the cloud. But both effects would modify 
the mid-infrared and 100 \um emission in the same way.
Therefore, variations of the dust size distribution or of emission
properties (hypothesis (3)) are more
directly traced by the mid-infrared to 100 \um emission ratio
than the mid-infrared emissivity per H atom. However it is only with the
\hi emission that we have access to the velocity information.

In field $A$, the emissivities in the LW2 and LW3 filters are enhanced by a factor $5\pm1$ and
$2.5\pm0.5$ in the filament with respect to the more diffuse atomic
gas. Since such a change is ruled out for large grain emission at 100 \ump,
we interpret the LW2 and LW3 emissivity variations as an evidence for 
variations in the small grains characteristics (abundance/emission
properties). 
Elements are still lacking to identify whether the emissivity variations
oberved here are due to variations in the abundance of the mid-infrared
emitters or to their emission properties. 
We recall that this filament has a very peculiar 
kinematic: a large velocity gradient across the filament which could
be a signature of a fast rotation around its long axis. 
We discuss below how
large velocity gradients might enhance the abundance of small dust grains.

Based on the mid-infrared to 100 \um ratio, in field $C$, 
the abundance of the particles
emitting in the LW2 and LW3 filters are found to be 
respectively $4.0\pm0.8$ and $2.1\pm0.4$ times
lower in the molecular region than in the atomic gas. A similar effect is also observed
with the IRAS 60 \um emission. 

\subsection{Turbulence and dust size distribution}

We discuss here a possible link between 
the  dust size distribution and the kinematics of the gas 
(i.e. the importance of the turbulent motions). 
The filament in field $A$, which has an aspect ratio of the order of 10 
\cite[]{miville-deschenes99b}, is
reminiscent of coherent structures of vorticity observed in turbulence
and thought to be responsible for the phenomenon of intermittency \cite[]{vincent94}.
\cite{falgarone95} investigated the decoupling of dust and gas 
in a turbulent flow as a function of grain size. They showed 
that the intermittency of turbulence enhances the decoupling of small 
dust grains from the gas motions 
because of the non-Gaussian distribution of the velocity gradients.
Thus the relative velocities 
of small dust particles can be significantly increased above that
reached
in a Kolmogorov turbulence. Qualitatively, the collision outcome is 
coagulation or grain fragmentation below and above some energy threshold. In these regions, 
the relative velocities might become sufficiently large that small dust grains are produced
by grain fragmentation in grain-grain collisions.
The high abundance of small dust particles in the filament of field $A$
could be a signature of such a fragmentation process. In the molecular gas, 
where the turbulent motions are smaller than in the \hi gas 
(Fig.~\ref{spectre_hi_co}), we may observe the reverse process: mutual coagulation of 
small grains could occur which would explain the drop of the
mid-infrared emissivity. 

The outcome  of a grain-grain collision (coagulation, bouncing or fragmentation)
depends on the kinetic energy of the colliding grains, on their respective masses
and on their material. Based on a numerical simulation of aggregate growth,
\cite{dominik97} have showed that two $\sim 0.1$ \um aggregates stick if their relative
velocity is $\delta v \lesssim 0.8$ \kmsp. But if it is higher than
$\sim 1$ \kmsp, the dust aggregates start to fragment.

Following \cite{draine85a}, the typical relative velocity of two grains of size $a_1$ 
in a Kolmogorov turbulent flow characterized by a velocity dispersion $v_{max}$ at a scale $l_{max}$
is given by the following equation:
\beq
v_{rel} = \left( \frac{v_{max}^{3/2}}{l_{max}^{1/2}} \right) \left( \frac{a_1 \rho_{gr}}{4n} \right)^{1/2}
\left( \frac{2 \pi}{\mu k T} \right)^{1/4},
\eeq
where $n$ and $T$ are the gas density and temperature and $\rho_{gr}$
is the grain mass density.
If we consider standard values for an \hi cloud like Ursa Major
($n=100$ \ccup, $T=200$ K) and Kolmogorov turbulence with 
$l_{max} =10$ pc and $v_{max}=5$ \kmsp,
the typical relative velocity for big grains 
($a_1 = 0.1$ \um and $\rho_{gr} = 3$ g cm$^{-3}$) 
would be $v_{rel} \approx 0.6$ \kmsp, which is close to the fragmentation
threshold established by \cite{dominik97}. But in field $A$, we observe
velocity gradients of 5 and 10 \kms pc$^{-1}$, which are larger
than the gradients characteristic of Kolmogorov turbulence at the
scale of the filament (0.3 pc). 
As shown by \cite{falgarone95}, these local large velocity gradients
may increase the relative velocity between small grains. 
Considering the strength of the velocity gradient observed here,
it is very likely that the relative velocity between large grains exceeds
the fragmentation threshold. 

In field $C$, the width of the CO line 
is $\sim 1.5$ \kms (see Fig.~\ref{spectre_hi_co}). If we consider that this
width is mostly due to turbulence and that the thickness of the Ursa Major cirrus is
$\sim 2$ pc, the typical velocity dispersion of 0.1 \um grains for the same
temperature and density as in field $A$ is $\sim 0.2$ \kms. This is lower than the
coagulation threshold determined by \cite{dominik97} and it could be
even lower since the density in the molecular region is higher than the mean
cirrus value.

This indicates that fragmentation and coagulation by 
grain-grain collisions in a turbulent flow may be at the origin of the abundance
variations of small dust grains observed in the Ursa Major cirrus. 
To estimate the timescales over which the
dust size distribution is significantly affected by this process, one should take
into account the collisions between grains of all sizes and, in the case of fragmentation, 
make some hypothesis on the size distribution of the fragments.
This is beyond the scope of the present paper.
\cite{draine85a} has estimated the depletion time scale of small grains 
in interstellar clouds to be $\sim 10^6$ years by considering only sticking
on large grains in Kolmogorov turbulence. Mutual coagulation of
small grains actually occurs faster
if we take intermittency into account \cite[]{falgarone95}.

\subsection{Variations of the $I_\nu({\rm LW2})/I_\nu({\rm LW3})$ ratio}

In addition to the spatial variations of the mid-infrared emissivity we have also observed
spatial variations of the $I_\nu({\rm LW2})/I_\nu({\rm LW3})$ ratio. For example, it is twice
larger in the \hi filament of field $A$ than in the diffuse gas that surrounds it. 
It is also observed to be twice larger in the atomic side of the \hip-H$_2$ transition
than in the molecular region. Similar variations of the $I_\nu({\rm LW2})/I_\nu({\rm LW3})$ ratio are
observed across photo-dissociation regions from
the photo-dissociated gas to the molecular gas \cite[]{abergel2000}. 

Using broad-band filters and CVF observations of L1630, \cite{abergel2000} showed
that the emission in the LW2 filter is correlated to the intensity of the 
7.7 \um aromatic band, and that the emission in the LW3 filter is correlated to the 
continuum level at 15 \ump. These CVF data show that
variations in the relative amplitude of the aromatic bands are minor and do 
not significantly affect the $I_\nu$(LW2)/$I_\nu$(LW3). 
It is unclear whether the aromatic bands and continuum at 15 \um
come from the same dust particles. 
Within the present understanding of 
the nature and emission properties of small dust grains it is
impossible to determine the origin
of the $I_\nu$(LW2)/$I_\nu$(LW3) ratio variations. 

\subsection{Coagulation of small grains}

Another example of the decrease of the small dust grain 
abundance in an \hip-H$_2$ interface is given by \cite{bernard99} who
characterize dust properties, in a section of the Polaris cirrus 
with moderate opacity  (A$_V\sim 1$), by combining IRAS and far-IR/sub-mm 
observations. In this cloud, there is no mid-IR emission, 
the 60/100 \um emission ratio is low and the large grain temperature, 
13 K, is lower than the mean cirrus value 17.5 K. These dust properties are 
characteristic of dense gas within molecular clouds. 
\cite{bernard99} explain these observations by the formation
of large fractal grains through the coagulation of small dust grains on large grains.
The decrease of the small dust grains abundance in the  \hip-H$_2$
interface of the Ursa Major cirrus could be an intermediate step in the dust evolution, 
between the diffuse \hi and denser environments.

\section{Conclusion}

\label{conclusion}

We have presented ISOCAM images (in two filters LW2 (5-8.5 \um) and
LW3 (12-18 \um)) 
of a high Galactic latitude cirrus transparent to
stellar photons. The 
angular resolution and brightness sensitivity of these images improve by one 
order of magnitude those provided by the 12 and 25 \um IRAS images. 
We have compared the ISOCAM images
with \hip, CO and $100 \mu m$ observations to measure the mid-IR emissivity (mid-IR emission from
small dust particles per hydrogen
atom) for distinct gas velocity components and across an interface between atomic and molecular gas.
We find significant variations in the mid-IR emissivity which we interpret
as changes in the abundance of small grains within the cloud.
The abundance of the small dust particles is larger by a factor 5 within a filament 
with a strong transverse velocity gradient, than in the surrounding diffuse \hip. 
We propose that this abundance enhancement results from the formation of small dust grains
through the fragmentation of larger grains in energetic grain-grain collisions 
induced by turbulence.
The abundance of small dust particles with respect to the larger grains 
is observed to be reduced (by a factor 4) in CO emitting molecular gas
with respect to its value in diffuse \hip. 
In this gas the turbulent motions have a smaller amplitude than
in the \hi gas. Within the proposed interpretation, we speculate that 
in such an environment grain/grain collisions are less energetic and they therefore 
lead to coagulation and not fragmentation. 

Spectroscopic observations
indicate that the LW2/LW3 color ratio measures the contrast
of the aromatic bands with respect to the underlying continuum emission.
A drop in the $I_\nu$(LW2)/$I_\nu$(LW3) ratio by a factor $\sim 2$ is observed from
the \hi filament to the diffuse \hi gas and from the \hi to the H$_2$ gas across the \hi-H$_2$
interface.  The LW2/LW3 color ratio is observed to decrease with the mid-IR emissivity. 
If the band contrast decreases for
increasing particles sizes, a plausible assumption in terms of solid state physics, 
the mid-IR emissivity and color variations could both be related to a change in the small-sizes
end of the dust size distribution.

\section*{Acknowledgements}

{\em The authors would like to thank Marc W. Pound for the very helpful CO observations
and the different teams at IAS, CEA-Saclay, ENS-Paris and ESA for their outstanding 
work and continuous support during all phases of the ISO project. The authors also thank the team at the DRAO
observatory for their help in the data reduction, Claire O'Neil for her help with the
writing and the referee, D. Hollenbach, for very helpful comments. 
The Fond FCAR du Qu\'ebec and the National Science and Engineering Research 
Council of Canada provided funds to support this research project.}

\end{document}